\documentclass[a4paper,11pt]{article}
\usepackage{graphicx, rotating,amsmath}
\usepackage{ifpdf}
\ifpdf
\usepackage{hyperref, pdfsync, epstopdf}	
\else
\usepackage[dvips,bookmarks]{hyperref}	
\fi
\hypersetup{colorlinks,bookmarksopen,bookmarksnumbered,citecolor=verdes,
linkcolor=blus,pdfstartview=FitH,urlcolor=rossos}
\def\hhref#1{\href{http://arxiv.org/abs/#1}{#1}} 
\newcommand{\eq}[1]{~{\rm (\ref{eq:#1})}}

\newcommand{\GeV}{\,{\rm GeV}}

\def\circa#1{\,\raise.3ex\hbox{$#1$\kern-.75em\lower1ex\hbox{$\sim$}}\,}

\newcommand{\NP}{Nucl. Phys.}

\newcommand{\PL}{Phys. Lett.}

\newcommand{\beq}{\begin{equation}}
\newcommand{\eeq}{\end{equation}}
\newcommand{\beqa}{\begin{eqnarray}}
\newcommand{\eeqa}{\end{eqnarray}}

\font\tenrsfs=rsfs10 at 11pt
\font\sevenrsfs=rsfs7
\font\fiversfs=rsfs5
\newfam\rsfsfam
\textfont\rsfsfam=\tenrsfs
\scriptfont\rsfsfam=\sevenrsfs
\scriptscriptfont\rsfsfam=\fiversfs
\def\mathscr#1{{\fam\rsfsfam\relax#1}}

\def\circa#1{\,\raise.3ex\hbox{$#1$\kern-.75em\lower1ex\hbox{$\sim$}}\,}
\makeatletter
\newcommand{\fig}[1]{~\ref{fig:#1}}
\def\art{\@ifnextchar[{\eart}{\oart}}
\def\eart[#1]#2#3#4#5#6{{\rm #2}, {#3 #4} {\rm (#6) #5} [{\hhref{#1}}]}
\def\hepart[#1]#2{{\rm #2, \hhref{#1}}}
\newcommand{\oart}[5]{{\rm #1}, {\em #2 \rm #3} {\rm (#5) #4}}

\newcounter{alphaequation}[equation]
\def\thealphaequation{\theequation\hbox to
0.6em{\hfil\alph{alphaequation}\hfil}}
\def\eqnsystem#1{
\def\@eqnnum{{\rm (\thealphaequation)}}
\def\@@eqncr{\let\@tempa\relax \ifcase\@eqcnt \def\@tempa{& & &} \or
  \def\@tempa{& &}\or \def\@tempa{&}\fi\@tempa
  \if@eqnsw\@eqnnum\refstepcounter{alphaequation}\fi
\global\@eqnswtrue\global\@eqcnt=0\cr}
\refstepcounter{equation} \let\@currentlabel\theequation \def\@tempb{#1}
\ifx\@tempb\empty\else\label{#1}\fi
\refstepcounter{alphaequation}
\let\@currentlabel\thealphaequation
\global\@eqnswtrue\global\@eqcnt=0 \tabskip\@centering\let\\=\@eqncr
$$\halign to \displaywidth\bgroup \@eqnsel\hskip\@centering
$\displaystyle\tabskip\z@{##}$&\global\@eqcnt\@ne
\hskip2\arraycolsep\hfil${##}$\hfil& \global\@eqcnt\tw@\hskip2\arraycolsep
$\displaystyle\tabskip\z@{##}$\hfil
\tabskip\@centering&\llap{##}\tabskip\z@\cr}
\def\endeqnsystem{\@@eqncr\egroup$$\global\@ignoretrue} \makeatother

\newcommand{\eV}{\,{\rm eV}}

\usepackage{multicol}
\usepackage{color}
\definecolor{rosso}{cmyk}{0,1,1,0.4}
\definecolor{rossos}{cmyk}{0,1,1,0.55}
\definecolor{rossoc}{cmyk}{0,1,1,0.2}
\definecolor{blu}{cmyk}{1,1,0,0.3}
\definecolor{blus}{cmyk}{1,1,0,0.6}
\definecolor{bluc}{cmyk}{1,1,0,0.1}
\definecolor{verde}{cmyk}{0.92,0,0.59,0.25}
\definecolor{verdec}{cmyk}{0.92,0,0.59,0.15}
\definecolor{verdes}{cmyk}{0.92,0,0.59,0.4}

\begin{document}
\begin{center}

\bigskip\bigskip

\color{black}
{\Huge\bf\color{rossos} The smallest neutrino mass}
\medskip
\bigskip\color{black}\vspace{0.5cm}

{
{\large\bf Sacha Davidson}$^a$,
{\large\bf Gino Isidori}$^b$,
{\large\bf Alessandro Strumia}$^c$.
}
\\[7mm]
{\it $^a$  CNRS/Universit\'e  Lyon 1,
IPN de Lyon,  Villeurbanne,  69622 cedex France}\\[3mm]
{\it $^b$ INFN, Laboratori Nazionali di  Frascati,
   Via E. Fermi 40, I-00044 Frascati, Italy   } \\ [3mm]
{\it $^c$ Dipartimento di Fisica dell'Universit{\`a} di Pisa and INFN, Italia}\\
\end{center}

\bigskip

\centerline{\large\bf\color{blus} Abstract}
\begin{quote}\large\color{blus}
We consider models where one Majorana neutrino is massless at tree level
(like the see saw with two right-handed neutrinos),
and compute the contribution to its
 mass $m$ generated by two-loop quantum corrections.
The result is  $m \sim 10^{-13}\eV$ in the SM and
 $m \sim 10^{-10} \eV \cdot (\tan\beta/10)^4 $ in the MSSM,
compatible with the restricted range suggested by Affleck-Dine baryogenesis.
\color{black}
\end{quote}

\bigskip

\section{Introduction}
Oscillation data~\cite{data} demand that two neutrinos are massive and strongly mixed;
in particular a roughly $\nu_\mu + \nu_\tau$ mass eigenstate is demanded by
the atmospheric anomaly.
The third neutrino mass eigenstate might be massless, and this possibility is
realized in various theoretical models, such as see-saw models 
with two right-handed neutrinos~\cite{2nu}.
Our study applies to generic models, where lepton number is broken at some high scale leaving
Majorana masses for two neutrinos.
Since $e,\mu,\tau$ have different Yukawa couplings, no symmetry demands that
the massless neutrino stays massless: with the inclusion of 
quantum corrections
all neutrinos become massive. In section~\ref{SM} 
we compute the neutrino mass generated by
renormalization-group equation (RGE) effects
in the Standard Model (i.e.~without new degrees of freedom up to the scale 
where neutrino masses are generated).
 In section~\ref{MSSM} we consider the Minimal Supersymmetric Standard Model (MSSM), and
in section~\ref{AD} we show that this quantum correction could allow successful
 Affleck-Dine (AD) baryogenesis via leptogenesis along the $LH_{\rm u}$ flat direction~\cite{AD}.


\section{Standard Model}\label{SM}
\begin{figure}[t]
$$\includegraphics[width=0.95\textwidth]{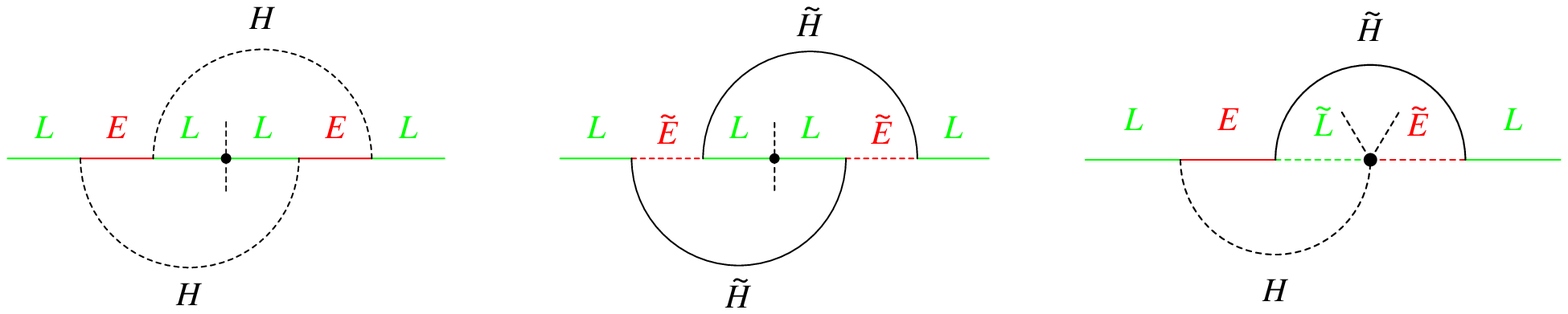}$$
\caption{\label{fig:Feyn2loop}\em Feynman diagrams (in the SU(2)-symmetric limit) that increase the rank of the neutrino mass matrix,
in the SM (left diagram) and in the MSSM.
The black dot denotes the effective operator generating 
the Majorana mass (in the MSSM case we omitted
diagrams proportional the $A$-term of $(LH_{\rm u})^2$).}
\end{figure}

Within the Standard Model (SM), Majorana neutrino masses are described by the effective operator $(L_i H)(L_j H)$ where $L$ and $H$ are the lepton and Higgs doublets.
Its coefficients can be parameterized by the neutrino mass matrix $m_{ij}$.
The dominant effect that increases the rank of $m_{ij}$ is the two-loop diagram 
shown in fig.~\ref{fig:Feyn2loop} (left). 
See~\cite{Babu} for earlier related studies.
The effect  is 
conveniently described in terms of the RGE
for $m$:
\begin{equation}
\label{eq:RGESM}
(4\pi)^2
\frac{d m}{d\ln  \mu} = m (\lambda  - 3 g_2^2+6\lambda_t^2) 
-\frac{3}{2} (m \cdot Y^T+
Y\cdot m )+
\frac{2}{(4\pi)^2} Y\cdot m\cdot Y^T+\cdots
\end{equation}
Here $m$ and $Y = \lambda_E^\dagger \cdot \lambda_E$ are $3\times 3$ matrices 
in flavour space ($\lambda_E$ is the charged-lepton Yukawa coupling), and 
$\lambda$, $g_2$ and $\lambda_t$ denote the Higgs self-coupling, the SU(2)$_L$
gauge coupling and the top-quark Yukawa coupling, respectively.
Notice that the flavor structure of the RGE is dictated by 
how $m$ and $\lambda_E$ transform under ${\rm U}(3)_L\otimes{\rm U}(3)_E$ 
flavor rotations of the left-handed lepton doublets ($L$) and singlets ($E$).
The first two terms on the r.h.s.~of eq.~(\ref{eq:RGESM}) arise at the 
one-loop level and have been computed in~\cite{SMRGE}: these terms do 
not change the rank of $m$.
From the explicit calculation of the first
two-loop diagram in  fig.~\ref{fig:Feyn2loop}
we have deduced the coefficient of the last term, which is the
dominant effect that increases the rank of $m$. 
The dots denote other effects at two-loop 
order and higher, that do not give qualitatively 
new effects.

We parameterize the neutrino mass matrix in the $e,\mu,\tau$ basis 
as $m = V^* {\rm diag}\,(m_1 ,m_2 ,m_3 ) V^\dagger$
where $m_1,m_2,m_3$ are complex eigenvalues,
$V =
R_{23}(\theta_{23}) \cdot
R_{13}(\theta_{13}) \cdot
\hbox{diag}\,(1,  e^{i \phi},1) \cdot
R_{12}(\theta_{12}) $,
and $R_{ij}(\theta_{ij})$ represents a
rotation by $\theta_{ij}$ in the $ij$ plane.
$\phi$ is the CP-violating phase in oscillations.
We denote the two leading eigenvalues of the neutrino mass matrix as
$m_{a}$ and $ e^{-2i\alpha} m_b$ with $m_{a}>m_{b}$,
and we denote as $e^{-2i\beta}m_c$ the smallest eigenvalue.
$\alpha$ and $\beta$ are the usual Majorana phases.
If neutrinos have normal hierarchy, then 
$a=3$ ($|m_3|=m_{\rm atm}\approx 0.05\eV$),
$b=2$ ($|m_2|= m_{\rm sun}\approx 0.009\eV$),
$c=1$ ($m_1 =  m_{\rm min}$).
If neutrinos have inverted hierarchy, 
one instead has $a=2$, $b=1$
(neglecting the solar mass splitting, the two heavier neutrinos are degenerate 
with mass $m_{\rm atm}$)
and $c=3$.
In both cases, $m_{\rm min}$ has a Majorana phase, $m_{\rm min} = |m_{\rm min}|e^{-2i\beta}$.
We can neglect the $\lambda_{e,\mu}$ couplings, such that $Y\simeq (0,0,\lambda_\tau^2)$.

In the `diagonalize and run' approach, eq.\eq{RGESM} 
can be converted into a RGE for the smallest eigenvalue $m_{\rm min}$:
\beq
\frac{dm_{\rm min}}{d\ln \mu}=
\frac{2\lambda_\tau^4}{(4\pi)^4}\bigg [(V_{\tau c} V_{\tau a}^*)^2 m_a +
(V_{\tau c} V_{\tau b}^*)^2  e^{-2i\alpha} m_b \bigg]+ \cdots
\eeq
where we only wrote the two-loop terms that generate it.

The `run and diagonalize' approach allows to write the explicit solution 
to eq.~\eq{RGESM} in the charged-lepton  eigenstate basis as:
\beq\label{eq:solm}
m(\mu) = r \begin{pmatrix} m^{(0)}_{ee} & m^{(0)}_{e\mu} & y m^{(0)}_{e\tau}\cr
m^{(0)}_{\mu e} & m^{(0)}_{\mu\mu} & y m^{(0)}_{e\tau}\cr
y m^{(0)}_{\tau e} & ym^{(0)}_{\tau\mu} & y^2z m^{(0)}_{\tau\tau}\end{pmatrix},\qquad
\eeq
where $m^{(0)}_{ij}$ are the initial values of the mass matrix (at some heavy scale where we assume
det$\, m^{(0)}=0$) 
and
\beq \ln r(\mu) = \int (\lambda - 3g_2^2+ 6 \lambda_t^2)dt~,
\qquad\label{eq:ySM}
\ln y(\mu)=-\frac{3}{2}\int \lambda_\tau^2 dt~,
\qquad
\ln z(\mu) =\frac{2}{(4\pi)^2} \int \lambda_\tau^4dt~,
\eeq
where $t= {\ln \mu}/{(4\pi)^2}$.
Eq.\eq{solm} shows that one loop effects generate a fake $m_{\rm min}$
if numerical inaccuracies or partial RGE resummation of higher orders terms
break the $y\cdot y\neq y^2$ relation, mimicking the effect of the two-loop term $z$.
Running down to the electroweak scale and 
computing the determinant, one gets the radiatively-generated 
light neutrino mass  and its Majorana phase:
\beq
m_{\rm min}  = (z-1) \bigg [(V_{\tau c} V_{\tau a}^*)^2 m_a +
(V_{\tau c} V_{\tau b}^*)^2  e^{-2i\alpha} m_b \bigg]~.
\eeq
Working at first order in $m_{\rm sun}\ll m_{\rm atm}$ 
and in $\theta_{13}\ll 1$,
and inserting numerical best-fit values $\theta_{\rm atm}=\pi/4$ and $\tan^2\theta_{\rm sun}=1/2$
in the subleading terms, one gets
%
%
%
\beq 
m_{\rm min} = (z- 1) 
 \bigg [ e^{2i \phi}
\frac{m_{\rm atm}}{4} \sin^2 2\theta_{\rm atm}\sin^2\theta_{\rm sun}
+  \frac{m_{\rm sun} e^{-2i\alpha} -3\sqrt{2} m_{\rm atm} \theta_{13}e^{i \phi}  }{18}\bigg]~,
\label{m1_norm}
\eeq
in the case of normal mass hierarchy, and
%
%
%
\beq
 m_{\rm min}  = (z- 1) e^{-2i(\phi+\alpha)} \bigg[\frac{m_{\rm atm}}{4} \sin^2 2\theta_{\rm atm}(\cos^2\theta_{\rm sun}+e^{2i \alpha }\sin^2\theta_{\rm sun} + \frac{2\sqrt{2}}{3} e^{i\phi}(e^{2i\alpha}-1)\theta_{13} )\bigg] 
\label{m1_inv}
\eeq
in the case of inverted mass hierarchy. 
We performed a 
global fit of present oscillation data\footnote{~$\theta_{12},\theta_{23},|\Delta m^2_{23}|,\Delta m^2_{12}$  
have been measured, there is an upper bound on $\theta_{13}$,
$\phi$ and $\alpha$ are unknown~\cite{data}.}
finding that the term in square brackets in eq.~(\ref{m1_norm})
lies between $1.4$ and 8 meV at $3\sigma$ confidence level.
The analogous term for inverted hierarchy in  eq.~(\ref{m1_inv})
lies between  $0$ and $16\,{\rm meV}$.
The RGE factor is 
\beq\label{eq:SUSY2}
z-1 \approx  \frac{2}{(4\pi)^4}\frac{m_\tau^4}{v^4} \ln \frac{M}{M_Z} \approx
0.85~10^{-12}\ln\frac{M}{M_Z} 
\eeq
where $v=174\GeV$; the numerical value, obtained from a numerical solution of SM RGE equations, 
agrees closely with the simple analytical approximation;
$M$ is the heavy scale where the initial condition 
det$(m^{(0)})=0$ holds. For $ M \circa{<} 10^{14}$~GeV 
we find $|m_{\rm min}| \sim 10^{-13}\eV$.

\medskip

In practice, no significant physical effects arises in the limit $m_{\rm min}\to 0$,\footnote{The SU(2)$_L$ analogous of the QCD $\theta$ angle gives anyway negligible
effects exponentially suppressed by $1/\alpha_2$.}
 so that
such a small value of $m_{\rm min}$ is not testable within the SM.
For example, oscillation predictions for $0\nu2\beta$~\cite{0nu2beta}
are the same for any $m_{\rm min}\ll m_{\rm sun}$.

\section{Minimal Supersymmetric Standard Model}\label{MSSM}
As shown in fig.\fig{Feyn2loop}, supersymmetry implies additional 
two-loop contributions to the neutrino mass matrix.
The diagrams in fig.\fig{Feyn2loop} are those surviving in the limit
of exact supersymmetry with $\mu=0$. In this limit their sums vanishes,
as dictated by the non renormalization theorem that only allows corrections to 
wave-functions and therefore forbids a RGE effect that increases the rank of the neutrino mass matrix.
Indeed RGE corrections have been computed up to two loop order~\cite{MSSMRGE} and the
$Y\cdot m \cdot Y$ term is absent.

However, supersymmetry must be broken, presumably at the weak scale.
Computing the diagrams in fig.\fig{Feyn2loop}, plus others with $A$-term vertices,
 one loses  the large RGE logarithm present in the SM
(the sum of the integrals is convergent) but gains a $\tan^4\beta$ enhancement,
because each one of the four $\tau$ Yukawa couplings $\lambda_\tau$
is enhanced by $\tan\beta$.
The induced neutrino masses can still be expressed 
by eqs.~(\ref{m1_norm})--(\ref{m1_inv}) with the replacement 
of the RGE factor with 
\beq (z - 1) \to  \frac{1}{(4\pi)^4} \frac{m_\tau^4}{v^4} \tan^4\beta \cdot f(m_{\tilde{L}}, 
m_{\tilde{E}}, m_H,m_{\tilde{H}})
\label{eq:SUSY}
\eeq
(in the large $\tan\beta$ limit),  where $v=174\GeV$
and $f$ is a finite adimensional 
order one function. The explicit form of  $f$ is not illuminating. 

\medskip

Furthermore, there is an additional effect \cite{GH} that gives
a potentially dominant contribution of relative order $g^2 \ln^2(M_{\rm max}/m_{\rm soft})/(4\pi)^2$
with respect to eq.\eq{SUSY}:
the  SUSY breaking  
term $A_{ij}(\tilde{L}_iH_{\rm u})(\tilde{L}_j H_{\rm u})$ 
 can  contribute to the neutrino mass matrix via
a gaugino-slepton loop.  Depending on taste
it can be either classified as a 1 or 2 or 3 loop effect.
We assume that the soft terms are flavor independent at
$M_{\rm max} = \min(M, M_{\rm med})$,
 where $M_{\rm med}$ is the mediation scale of soft terms, equal
 to $M_{\rm Pl}$
 in supergravity-mediated models.
Slepton masses get corrected by flavor-dependent RGE effects;
the eigenvectors of  $A_{ij}$ get rotated relative to those of $m_{ij}$
and the rank of the $A_{ij}$-term matrix is increased already
by one loop RGE-running between $m_{\rm soft}$ and
$M_{\rm max}$. 
 Indeed, the one loop RGE for $\hat{A}_{ij}\equiv v^2 A_{ij}/m_{ij}$ is
\beq
(4\pi)^2 \frac{d\hat{A}_{ij}}{d\ln\mu} =2 (\delta_{i\tau} +\delta_{j\tau}) \hat{A}_\tau \lambda_\tau^2+\cdots
\eeq
where $A_\tau=\hat{A}_\tau \lambda_\tau$ is the $A$-term of the $\tau$-Yukawa coupling and
$\cdots$ denotes other terms not crucial for the present discussion.
The solution has the form
\beq\label{eq:solSUSY}
\hat{A}(\mu) =  \begin{pmatrix} \hat{A}^{(0)} & \hat{A}^{(0)} &  
\hat{A}^{(0)} + \epsilon \hat{A}_\tau \cr
\hat{A}^{(0)} & \hat{A}^{(0)} & \hat{A}^{(0)} + \epsilon \hat{A}_\tau \cr
\hat{A}^{(0)} + \epsilon \hat{A}_\tau & 
\hat{A}^{(0)} + \epsilon \hat{A}_\tau & \hat{A}^{(0)}+ 2\epsilon \hat{A}_\tau \end{pmatrix},\qquad
\epsilon \simeq \frac{\lambda_\tau^2}{(4\pi)^2} \ln\frac{M_{\rm max}}{\mu}\ .
\eeq
If $A^0_{ij}=\hat{A}^{(0)} m_{ij}^{(0)}/v^2$ has one zero eigenvalue,
the additive correction in eq.\eq{solSUSY}
transforms it into a small ${\cal O}(\epsilon^2) \hat{A}_\tau$ eigenvalue in
$A_{ij}$, 
justifying our above estimate.\footnote{We here elaborate on the possibly surprising claim that
one-loop RGE running generates no ${\cal O}(\epsilon)$ eigenvalue and generates
a two-loop-like ${\cal O}(\epsilon^2)$ eigenvalue.
Notice that the structure of the additive terms to $A_{ij}$ in eq.\eq{solSUSY} 
(0 or 1 or 2 depending on how many $\tau$ there are in $ij$) is exact, 
such that we do control ${\cal O}(\epsilon^2)$ terms.
Furthermore, our result is compatible with the general statement~\cite{Kazakov}
that RGE effects in softly broken supersymmetry 
can be condensed into a  renormalization of the superfields,
with renormalization factors and couplings appropriately promoted to spurion superfields.
In our case it (roughly) means
that at one loop  $\tilde{m}_{ij}\equiv m_{ij} + \theta\theta A_{ij}+\cdots$ only gets corrected via a
$\tilde{y} \approx  y + \theta\theta \epsilon A_\tau+\cdots$ multiplicative
renormalization of the $L_\tau$ superfield, where $y$ is MSSM analogous of
SM $y$ in~\eq{ySM}.
The vanishing of $\det \tilde{m}$ implies the vanishing of $\det m$
and of a combination of $A\times m$ (verified by eq.\eq{solSUSY}), and
does not imply the vanishing of $\det A$.}
Due to the large uncertainty (all sparticle masses are unknown),
we  just estimate  the slepton-gaugino loop~\cite{GH} contribution to be: 
$$m_{\rm min}\sim  m_{\rm atm} 
\frac{g^2}{64 \pi^2} \frac{\lambda_\tau^4}{(4\pi)^4} \frac{m_\chi \hat{A} }{m^2_{\rm soft}}
\ln^2\frac{M_{\rm max}}{m_{\rm soft}}
\sim  10^{-10}\eV\cdot (\frac{\tan\beta}{10})^4,$$
comparable to the 2 loop contribution of eq.\eq{SUSY}.

%
%


\section{Affleck-Dine leptogenesis}\label{AD}
In the supersymmetric context, such a small neutrino mass
can have phenomenological consequences.
Indeed, recent analyses found that a scalar condensate along the $LH_{\rm u}$
flat direction
can produce the observed baryon asymmetry if $m_{\rm min}\sim 10^{-(12\div 9)}\eV$~\cite{AD}, where
the uncertainty is due to our lack of knowledge about the reheating temperature, sparticle masses
and CP phases.

The results of~\cite{AD} cannot be immediately applied to our scenario 
because their neutrino masses are not   radiatively generated.
Nevertheless, let us first summarize some of the key points of~\cite{AD}.
As usual, the $B-L$ conserving sphalerons transfer a lepton asymmetry into a baryon asymmetry:
this singles out the $LH_{\rm u}$ direction.
It develops, during inflation, a large scalar condensate
$\varphi \equiv \langle \tilde{L} \rangle \simeq \langle H_{\rm u}\rangle$
if $\varphi $  has a soft mass$^2$  of order $- H^2$
from the inflationary vacuum energy, where $H$ is the expansion rate during inflation, then
one expects $\varphi \sim  v\sqrt{H/m_{\rm min}}$.
The smallest neutrino mass $m_{\rm min}$ gives the dominant effect because it allows the largest $\varphi$.
When $H^2$ decreases below the  mass$^2$ terms in $V(\varphi)$
(the  normal soft masses, $m^2_{\rm soft}$, 
plus  thermal corrections of ${\cal O}(T^2)$, relevant if the reheating temperature after inflation is high enough)
the condensate starts to oscillate generating a sufficient lepton asymmetry,\footnote{Unless
the system  remains trapped in one of the unphysical vacua often present in the MSSM; thermal effects allow to partially predict which local minimum is dynamically selected as the vacuum~\cite{CCB}.}
because the potential contains a $\varphi^4$ term that breaks lepton number,
 coming from the $A$-term of the $(LH_{\rm u})^2$ operator.

Let us now come to the case of radiatively-generated $m_{\rm min}$.
The dynamics of AD leptogenesis is not directly controlled
by   neutrino masses, but
by the $\varphi^4$ and $|\varphi|^6$ terms in the $V(\varphi)$ potential at the beginning of oscillations,
respectively generated by the ${A}$-term
and by the $F$-term of the neutrino mass operator $(LH_{\rm u})^2$. 
The $\varphi^4$ term acts as the source of lepton-number breaking
and the $\varphi^6$ term limits the initial vev of $\varphi$.
Denoting by $r_4$ and $r_6$ the correction factors 
of these terms with respect to
the `standard' values considered in earlier analyses~\cite{AD},
the final amount of baryon asymmetry  gets corrected by $r_4/r_6$.


We therefore need to compute $r_4$ and $r_6$.
For simplicity we assume that $\varphi$ remains below the scale 
of new physics that generates the $LH_{\rm u}$ operator.\footnote{In the see-saw scenario, 
this new physics are right-handed neutrinos of mass $M$, 
with $M < 10^{14\div 15}\GeV$ if we want to remain in a perturbative regime. 
It is not clear to us what happens if instead $\varphi> M$;
possibly $\varphi$ would slide up to the GUT scale (around $10^{16}\,{\rm GeV}$)
rather than being limited by the $|\varphi|^6$ term,
giving rise to a dynamics somewhat different from the one studied in~\cite{AD}.}
The previous section suggests that, similarly to the lightest neutrino mass,
the $\varphi^4$ and the $|\varphi|^6$ terms in $V(\varphi)$ 
are generated by quantum corrections, such 
that today (after the end of inflation, at temperature $T\ll m_{\rm soft}$)
$r_{4,6}$ are not much different from one. 
However, quantum corrections depend on sparticle masses,
which had different values during the epoch relevant for AD leptogenesis.
There are various effects.
Corrections to soft terms of order $H^2$ (inflationary masses) and of order
$T^2$ (thermal masses) do not qualitatively change our results, because
they generically break supersymmetry.
The time dependence of $\varphi$ provides one more source of SUSY-breaking
via the $D$-terms; furthermore, $\varphi(t)$ directly contributes to $V(\varphi)$
when inserted into  higher dimensional $D$-terms such as $(L  \partial  H_{\rm u})^2/M^3$.
On the contrary, the large vev $\varphi$, inserted in the $\lambda_\tau LEH_{\rm d}$ 
coupling, generates a  large supersymmetric mass $\sim \lambda_\tau \varphi$ 
for the $E$ and $H_{\rm d}^-$ particles and sparticles. 
This suggests that  $r_4,r_6  \sim m^2/(\lambda_\tau\varphi)^2$
where $m^2$ are SUSY-breaking masses coming from the effects discussed above.
Then $r_4/r_6$ remains of order one, such that
AD leptogenesis remains successful for the 
standard value $m_{\rm min}\sim 10^{-9\div 12}\eV$.
We have shown that a value in this range does not need contrived flavor models
and can be naturally generated by quantum corrections.
This encouraging result might be tested  in a more stringent way if
sparticles will be discovered, 
and if their masses and especially $\tan\beta$ will be measured.



\section{Conclusions}
Assuming that two neutrinos have Majorana masses and that
the lightest neutrino is massless at tree level, we computed
the mass generated by quantum corrections, and its Majorana phase.
In the SM two loop RGE running
gives $|m_{\rm min}|\sim 10^{-13}\eV$.
In the MSSM supersymmetry breaking generates various 
flavor matrices that contribute in different ways; the typical
result is $|m_{\rm min}|\sim 10^{-10}\eV(\tan\beta/10)^4$, enhanced
by four powers of $\tan\beta$.
Such a small neutrino mass is compatible with the restricted range of values that
 allows successful Affleck-Dine leptogenesis 
along the $LH_{\rm u}$ flat direction.


\paragraph{Acknowledgements}
A.S.\ thanks L. Boubekeur, R. Rattazzi for useful comments and P. Ciafaloni and A. Romanino for an old collaboration
where 2 loops tools employed in the present paper were developed.
We thank K.~Babu and E.~Ma for drawing our attention to their earlier study~\cite{Babu}.


\end{document}